\documentclass[12pt]{article}

\input epsf
\def\beq{\begin{eqnarray}}
\def\eeq{\end{eqnarray}}

\def\lsim{\mathrel{\rlap{\lower3pt\hbox{\hskip0pt$\sim$}}
     \raise1pt\hbox{$<$}}}         
\def\gsim{\mathrel{\rlap{\lower4pt\hbox{\hskip1pt$\sim$}}
     \raise1pt\hbox{$>$}}}         

\newcommand{\comment}[1]{}
\usepackage{amsmath}
\usepackage{amsfonts}
\usepackage{verbatim}
\usepackage{graphicx}
\usepackage{subfigure}
\usepackage{amssymb}
\usepackage[T1]{fontenc}
\usepackage{calligra}

\renewcommand{\comment}[1]{}

\begin{document}

\begin{titlepage}

\thispagestyle{empty}

\begin{flushright}
{NYU-TH-10/10/37}
\end{flushright}
\vskip 0.9cm

\centerline{\Large \bf Unitarity Check in Gravitational Higgs Mechanism}
\vskip 0.2cm
\centerline{\Large \bf }                    

\vskip 0.7cm
\centerline{\large Lasha Berezhiani and Mehrdad Mirbabayi}
\vskip 0.3cm
\centerline{\em Center for Cosmology and Particle Physics, Department of Physics,}
\centerline{\em New York University, New York, 
NY  10003, USA}

\vskip 1.9cm

\begin{abstract}

The effective field theory of massive gravity had long been formulated  in a generally covariant way arXiv:hep-th/0210184.  Using this formalism, it  has been found recently that there exists a class of  massive nonlinear theories that are free of the Boulware-Deser ghosts, at least in the decoupling limit	arXiv:1007.0443. In this work we study other  recently proposed models that go under the name of "gravitational Higgs theories" arXiv:1002.3877, 	arXiv:1008.5132. We show that these  models, although seemingly different from the effective field theories of massive gravity, are in fact equivalent to them. Furthermore, based on the results obtained in the effective field theory approach, we conclude that the gravitational Higgs theories need the same adjustment of the Lagrangian  to avoid the ghosts. We also show the equivalence between the noncovariant mode decomposition used in the Higgs theories, and the covariant St\"uckelbergization adopted  in the effective field theories, thus, proving that the presence or absence of the ghost is independent of the parametrization used in either theory.

\end{abstract}

\vspace{3cm}

\end{titlepage}

\newpage

In order for the theory of a massive spin-2 particle to be consistent with four-dimentional Poincar\'e symmetry, it should propagate five physical degrees of freedom: helicity-$\pm 2$, helicity-$\pm 1$ and helicity-$0$. The only ghost-free and tachyon-free quadratic potential that describes these modes is that of Fierz and Pauli \cite{FP,Nieu}.

As it is well known in the Fierz-Pauli theory, even in zero-mass limit, the helicity-0 mode couples to the trace of the matter energy-momentum tensor with the same strength as the helicity-2 does \cite{vDVZ}, causing inconsistency with current observations. However, it has been argued by Vainshtein \cite{Arkady} that this discontinuity can be removed by non-linear effects as a result of the screening of the helicity-0 mode at observable scales, which on the other hand would make the theory compatible with the known empirical data \cite{DDGV}. But because of the same non-linearities the theory is afflicted with ghost \cite{BD}, sometimes called the Boulware-Deser mode. Although it is infinitely heavy in a Minkowski space, it emerges as a light sixth degree of freedom on any locally nontrivial background \cite{Creminelli,DeffayetRombouts,Andrei}.

The easiest way to see this is to proceed in analogy with a massive non-Abelian vector field \cite{Khriplovich}. In massive gravity, after taking the decoupling limit\footnote{This limit corresponds to the $m \ll E < \Lambda_5$ energy range.} \cite{AGS}
\beq
\label{Lambda5}
m\rightarrow 0,\quad M_{pl}\rightarrow \infty, \quad \Lambda _5 \equiv (M_{pl} m^4)^{1/5} - \text{fixed},
\eeq
with $m$ the mass of graviton, the helicity-0 mode, $\varphi$, decouples from the rest of the constituents of the physical graviton and is described by the following schematic Lagrangian
\beq
\mathcal{L}_\varphi=\frac{3}{2} \varphi \Box \varphi+\frac{(\partial^2 \varphi)^3}{\Lambda _5 ^5}.
\label{Lpi}
\eeq
To see the existence of more than one degree of freedom in the theory given by \eqref{Lpi}, one could notice that there appear fourth derivatives of $\varphi$ in equations of motion, meaning that the two initial conditions are not enough for the Cauchy problem to be well-defined \cite{DeffayetRombouts}. Moreover, on a locally nontrivial background with $\Box \varphi \propto T \neq 0$, the cubic interaction could generate the four derivative quadratic term $(\partial^2 \varphi)^2$ in effective action. This will lead to the existence of a ghost which is light enough for making the theory ill-defined before it reaches the strong coupling regime \cite{Creminelli}.

However recently in \cite{giga} it was realized that the particular completions of Fierz-Pauli that have the highest possible high energy cutoff ($\Lambda_3 \equiv (M_{pl}m^2)^{1/3}$) are ghost-free at least in the decoupling limit (when $\Lambda_3$ is fixed and $M_{pl}\to \infty$ and $m\to 0$). This provides us with a necessary condition that must be satisfied by any theory of massive gravity to be stable.

It is therefore reasonable to ask whether a given model fulfills this criterion, and we try to answer this question for the rather attractive candidate of mass generation through Higgs mechanism \cite{'tHooft,Kakushadze,slava2,slava3}. The crucial observation is that while a homogeneous Higgs condensate can only give rise to cosmological constant, a coordinate dependent one generates different non-derivative graviton self-couplings via the Higgs kinetic term that necessarily involves metric. In particular \cite{Kakushadze} and \cite{slava2} have proposed models which reproduce the correct Fierz-Pauli mass term up to quadratic order, and in \cite{slava3} possible extensions of \cite{slava2} to get lower Vainshtein radius \cite{Arkady} has been studied.

We first briefly review the effective theory of massive gravity \cite{AGS} and the findings of \cite{giga}. Defining
\beq
\label{hmn}
H_{\mu \nu}=g_{\mu \nu}-\partial_ \mu \phi^A \partial_ \nu \phi^B \eta _{AB},
\eeq
where $A,B=0,1,2,3, ~\eta_{AB}=\operatorname{diag}(-1,1,1,1)$ and $\phi^A$ transform as scalar fields under the general coordinate transformations, a gauge invariant Lagrangian for massive gravity can be constructed in the following way
\beq
\mathcal{L}=M_{pl}^2\sqrt{-g}R-\frac{m^2 M_{pl}^2}{4} \sqrt{-g}~V\left(g^{\mu\nu}, H_{\mu \nu}\right)\,.
\label{L}
\eeq
Here $V(g^{\mu\nu}, H_{\mu \nu})$ is a polynomial in $H_{\mu \nu}$ and up to the cubic order is given by
\beq
V(H_{\mu \nu})=H ^2 _{\mu\nu}-H^2+c_1 H_{\mu \nu}^3
              +c_2 H H_{\mu \nu}^2+c_3 H^3+\operatorname{O}(H^4),
\label{pol}
\eeq
with all indices contracted using $g^{\mu\nu}$. This theory possesses the following background solution
\beq
g_{\mu\nu}=\eta_{\mu\nu},\qquad \phi^A=\delta_{\mu}^A x^\mu.
\label{vac}
\eeq
Considering perturbations $h_{\mu\nu}\equiv g_{\mu\nu}-\eta_{\mu\nu}$ and $\pi^\alpha\equiv x^\alpha-\phi^\alpha$ one obtains 
\beq
\label{eq:H_pi}
H_{\mu\nu}=h_{\mu\nu}+\partial_\mu\pi_\nu+\partial_\nu\pi_\mu
          -\partial_\mu\pi_\alpha\partial_\nu\pi^\alpha\,,
\eeq
with $\pi_\alpha=\eta_{\alpha\beta}\pi^\beta$. Note that in the unitary gauge $\pi_\alpha$ is set to zero, so that $H_{\mu\nu}=h_{\mu\nu}$, and the potential \eqref{pol} gives the correct ghost-free quadratic Fierz-Pauli term.

The dynamics of helicity-0 and helicity-1 modes can be extracted by making the decomposition
\beq
\label{eq:parametrization}
\pi_\alpha=A_\alpha+\partial_\alpha\varphi\,,
\eeq
in $H_{\mu\nu}$. Doing so in pure Fierz-Pauli action, $m^2M_{pl}^2(H^2-H_{\mu\nu}^2)/4$, one can see that the kinetic term for $\varphi$ is obtained from the mixing $\varphi (\eta_{\mu\nu}\Box-\partial_\mu\partial_\nu)h^{\mu\nu}$. After diagonalization, canonical normalization $\varphi^c\equiv\Lambda_3^3\varphi$, and keeping only the most strongly coupled interactions, the Lagrangian for $\varphi$ becomes \cite{AGS}
\beq
\label{eq:gravity_decouple}
{\cal L}_\varphi=-\frac{1}{2}(\partial_\mu\varphi^c)^2
          +\frac{1}{2\Lambda_5^5}[(\Box \varphi^c)^3
	            -\Box\varphi^c(\partial_\mu\partial_\nu\varphi^c)^2]\,.
\eeq
The interactions of the form $(\partial^2 \varphi)^3$, however, result in a ghost \cite{Creminelli,DeffayetRombouts}.

The observation of \cite{giga} was that if one tunes the coefficients in the expansion of $V$ to all orders, so as to push the cutoff to $\Lambda_3\equiv (M_{pl}m^2)^{1/3}$ then

i) The dangerous terms $(\partial^2 \varphi)^n$, that may give rise to a ghost on a local background vanish from the Lagrangian, up to total derivatives.

ii) In the new decoupling limit:
\beq
m\to 0,\quad M_{pl}\to \infty, \quad \Lambda _3 - \text{fixed},
\label{decoupling}
\eeq
the Bianchi's identities continue to hold. That is, the terms which mix helicity-0 and helicity-2, $h^{\mu \nu} X_{\mu \nu}(\varphi)$, satisfy the transversality condition $\partial^\mu X_{\mu\nu}=0$. Here $X_{\mu\nu}$ is a symmetric tensor which is a function of the longitudinal degree of freedom and is given in \cite{giga,dato}.

These two points guarantee the absence of the ghost in the decoupling limit \eqref{decoupling}. For instance the fine-tuned cubic coefficients selected in this way are\footnotemark
\beq
c_1=2c_3+\frac{1}{2}, \qquad c_2=-3c_3-\frac{1}{2}.
\label{coef3}
\eeq

\footnotetext{It is worth mentioning that one particular set of coefficients corresponding to $c_3=1/4$ has been obtained in \cite{cubic}, in the framework of auxiliary extra dimension of \cite{GG,Claudia}.}

\vskip 0.3cm

\subsubsection*{Unitarity Check}

We use this result to analyze the models of \cite{Kakushadze} and \cite{slava2,slava3}. To this end the unitary gauge, in which all auxiliary fields have been absorbed inside the metric perturbations, provides the best framework since it unifies all different possible ways of introducing scalars (or pions in the language of effective theory \cite{AGS}).

In \cite{slava2} the dynamical generation of the graviton mass term is achieved by adding four scalar fields $\phi ^A,~A=0,1,2,3,$ with high-derivative interaction terms to general relativity. These terms are considered to be a function of the following field space tensor
\beq
H^{AB}=g^{\mu \nu} \partial_\mu \phi^A \partial_\nu \phi^B,
\label{hab}
\eeq
with field space indices being raised and lowered by $\eta_{AB}=\text{diag}(-1,1,1,1)$. The Lagrangian is then given by
\beq
\mathcal{L}=M_{pl}^2\sqrt{-g}R+\frac{m^2 M_{pl}^2}{4} \sqrt{-g}~V\left(H^{AB}\right)\,,  \\
\label{lag}
V\left(H^{AB}\right)=3 \left( \left( \frac{1}{4} H\right)^2-1 \right)^2-\tilde{H}^A _B \tilde{H}^B _A\,,
\label{pot}
\eeq
where $H \equiv \eta_{AB}H^{AB}$, and $\tilde{H}^A _B \equiv H^A _B-\frac{1}{4} \delta _B ^A H$ denotes the traceless part of $H^{AB}$. The background solution of the equations of motion which corresponds to the Minkowski space is given by \eqref{vac}.

For further analysis it is useful to rewrite the Lagrangian in terms of a new variable $\bar{h}^{AB}\equiv H^{AB}-\eta^{AB}$. The latter field redefinition is useful because it vanishes on vacuum (\ref{vac}), thus making it easy to truncate the expansion of the potential at desired order. The expression for the potential (\ref{pot}) in terms of the new variable reads
\beq
V=\left( \bar{h}^2-\bar{h}^A _B \bar{h}^B _A \right)+\frac{3}{4^2}\bar{h}^3+\frac{3}{4^4}\bar{h}^4.
\label{potbar}
\eeq
And in general, any $V(H^{AB})$ can be expanded in terms of products of monomials of the form
\beq
\label{eq:monomial}
\bar{h}^{A_1 B_1} \ldots \bar{h}^{A_n B_n}\eta_{B_n A_1}\ldots \eta_{B_{n-1} A_n}\,.
\eeq
But in unitary gauge, \eqref{hab} implies that $\bar{h}^{AB}=g^{AB}-\eta^{AB}$ so that
\beq
\eta_{AC}\bar{h}^{CB}=\eta_{AC}(g^{CB}-\eta^{CB})=-g^{CB}(g_{CA}-\eta_{CA})
               = -g^{CB}h_{CA}=-g^{CB}\delta_C ^\mu \delta_A ^\nu H_{\mu \nu}\,,
\eeq
with $H_{\mu\nu}$ defined in \eqref{hmn}. Notice that the first and last equalities hold only in unitary gauge. Therefore in this gauge
\beq
\bar{h}^{A_1 B_1} \ldots \bar{h}^{A_n B_n}\eta_{B_n A_1}\ldots \eta_{B_{n-1} A_n}
=(-1)^n H_{\mu_1 \nu_1} \ldots H_{\mu_n \nu_n} g^{\nu_n \mu_1} 
                    \ldots g^{\nu_{n-1} \mu_n}
\label{sim}
\eeq
and any potential written in terms of $\bar{h}^{AB}$ can readily be translated in terms of $H_{\mu\nu}$ and its coefficients be compared to \eqref{pol}. In particular \eqref{potbar} propagates ghosts beyond quadratic order, because it does not coincide with \eqref{pol} for any value of $c_3$, after taking into account \eqref{coef3}. In the appendix we will show that \eqref{sim} holds in arbitrary gauge which means that the scalar fields introduced to restore diffeomorphism invariance are closely related in two theories.

In \cite{slava3} the problem of constructing potentials $V(H^{AB})$ with smaller Vainshtein radius ($R_V$) around a static source of mass $M_0$, (or equivalently larger high energy cut-off) has been studied. It was observed that while in the original model $R_V = (M_0/M_{pl}^2m^4)^{1/5}$ (corresponding to a cut-off equal to $\Lambda_5=(M_{pl}m^4)^{1/5}$), it can be lowered by order by order adjustment of terms in the perturbative expansion of $V(\bar{h}^{AB})$,  until the asymptotic value of $R_V=(M_0/M_{pl}^2m^2)^{1/3}$ (cut-off equal to $\Lambda_3=(M_{pl}m^2)^{1/3}$) is reached. However according to \cite{giga} the only completion of Fierz-Pauli that is potentially ghost-free is the one with the cut-off pushed to the highest possible value, namely $\Lambda_3$. Therefore we expect all completions of $V(\bar{h}^{AB})$ except the very last one with $\Lambda_3$ cut-off  to suffer from ghosts.

\vskip 0.5 cm

The massive gravity proposed in \cite{Kakushadze} is also based on high derivative kinetic terms for a set of scalar fields $\phi ^A,~A=0,1,2,3$. With a little change of notation and restriction to $4D$, the Lagrangian is considered to be given by
\beq
\label{eq:KakushAction}
&S_Y=\int d^4x \sqrt{-g}\left(M_{pl}^2 R-\frac{3}{2}M_{pl}^2 m^2 V_Y(Y,U/\sqrt{-g})\right)\,,&\\
&Y\equiv g^{\mu\nu}Y_{\mu\nu}\equiv 
           g^{\mu\nu}\partial_\mu \phi^A \partial_\nu \phi^B\eta_{AB}\,,&\\
&U\equiv \frac{1}{4!}\epsilon^{\mu\nu\rho\sigma}
           \partial_\mu \phi^A \partial_\nu \phi^B \partial_{\rho}\phi^C \partial_\sigma \phi^D
             \epsilon_{ABCD}=\sqrt{-\operatorname{det}(Y_{\mu\nu})}\,.&
\eeq
The potential $V_Y$ is chosen in a way to yield the background solution \eqref{vac}.

This theory may be compared to \eqref{L} in unitary gauge in which the perturbations of $\phi$ fields vanish and $Y_{\mu\nu}=\eta_{\mu\nu}$. To this end it is necessary to expand $\sqrt{-g}$ as well as $V_Y$ in terms of metric perturbations since the latter, in general, contains constant and linear terms in $h_{\mu\nu}$. As an illustration we consider the simplest possible potential that gives rise to Fierz-Pauli on a Minkowski background at quadratic level, namely $V_Y=\Lambda+Y+\lambda Y^2$ with $\Lambda=-2$ and $\lambda=-1/12$. After proper normalization, to cubic order 
\beq
\label{eq:V_Y_3}
6\sqrt{-g}V_Y=4+h_{\mu\nu}^2-h^2-\frac{4}{3}h_{\mu\nu}^3+2 h h_{\mu\nu}^2
              -\frac{5}{12}h^3+\operatorname{O}(h^4)\,,
\eeq
while \eqref{L},\eqref{pol} and \eqref{coef3} lead to 
\beq 
\label{eq:V_eff_3}
\sqrt{-g}V=h_{\mu\nu}^2-h^2+ \left(2c_3-\frac{3}{2}\right)h_{\mu\nu}^3-\left(3c_3-2\right) h h_{\mu\nu}^2
              +\left(c_3-\frac{1}{2} \right)h^3+\operatorname{O}(h^4)\,,
\eeq
where all contractions have been done with $\eta^{\mu\nu}$ (note that $g^{\mu\nu}$ had been used for contraction in \eqref{pol}). Equations \eqref{eq:V_Y_3} and \eqref{eq:V_eff_3} do not coincide for any value of $c_3$, therefore the dynamics of the helicity-0 mode in the decoupling limit is schematically given by \eqref{Lpi}, resulting in a ghost at cubic level. Note that the inclusion of higher order terms in the expansion \eqref{eq:V_Y_3} cannot cure the instability since their contributions to \eqref{Lpi} are suppressed by higher scales\comment{ and vanish in this limit}. Nevertheless it is in principle possible to construct $V_Y$ such that it reproduces the expansion \eqref{eq:V_eff_3}.

\vskip 0.3 cm

We would also like to make a general comment regarding Higgs mechanism. If the mechanism is indeed higgs-like there must be heavy degree(s) of freedom (higgs bosons) that unitarize amplitudes at high energy. In other words, with increasing energy, operators with dimension $>4$ become more and more important and eventually the theory of massive gravity becomes strongly coupled, unless the higgs boson starts to contribute and keeps the theory perturbative. Otherwise graviton ceases to exist as an asymptotic degree of freedom. The models shown to have ghost are similar to the Fierz-Pauli gravity in the sense that the only candidate for the abovementioned heavy degree of freedom is a ghost \cite{DeffayetRombouts}. On the other hand the procedure of reducing the Vainshtein scale, outlined in \cite{slava3} tends to remove the ghost order by order in the decoupling limit, leaving only five degrees of freedom which become strongly coupled in the vicinity of $\Lambda_3$. Hence the models discussed should be considered as effective field theoretic descriptions of massive gravity rather than the Higgs mechanism for it.

\subsubsection*{Is the Ghost a Result of Bad Parametrization?}

One may wonder whether the ghost appearing in theories of massive gravity with non-tuned coefficients is an artifact of the St\"uckelberg parametrization \eqref{eq:parametrization}, since it contains time derivatives. Here we show, using another parametrization of $\pi_\alpha$ without time derivatives, that once auxiliary fields are integrated out the presence of ghost becomes evident and therefore it is not a byproduct of parametrization. As an illustration first consider the Lagrangian of a massive vector field amended by a new quadratic term
\beq
\label{eq:massiv_vector}
{\cal L}=-\frac{1}{4}F_{\mu\nu}^2-\frac{1}{2}m^2A_\mu^2
          +\frac{1}{2}\alpha (\partial_\mu A^\mu)^2\,.
\eeq
Substituting $A_\mu=a_\mu+\partial_\mu\varphi$, the last term gives rise to a high derivative kinetic term for $\varphi$. In particular after canonical normalization, $\varphi^c\equiv m\varphi$, and taking the decoupling limit $m,\alpha \to 0$ with $\alpha /m^2=const$ one obtains
\beq
\label{eq:mass_vector_decouple}
{\cal L}_{decouple}=-\frac{1}{4}F_{\mu\nu}(a_\mu)^2
           -\frac{1}{2}(\partial_\mu\varphi)^2
           +\frac{1}{2}\frac{\alpha}{m^2} (\Box \varphi)^2\,,
\eeq
which clearly describes four degrees of freedom, one of them being a ghost.

On the other hand consider the parametrization $A_0=\chi$ and $A_i=a_i^T+\partial_i\tilde\varphi$ with $\partial_ia^T_i=0$. Inserting this into \eqref{eq:massiv_vector}, the Lagrangian for the scalar fields, which decouple from $a^T_i$ due to transversality of the vector mode, becomes
\beq
{\cal L}_{scalar}=\frac{1}{2}(\partial_i\dot{\tilde\varphi}-\partial_i\chi)^2
          + \frac{1}{2}m^2[\chi^2-(\partial_i\tilde\varphi)^2]
          + \frac{1}{2}\alpha(\dot\chi-\Delta\tilde\varphi)^2\,.
\eeq
However after integrating out $\chi$ and taking the decoupling limit this reduces to the second and third terms on the r.h.s. of \eqref{eq:mass_vector_decouple}. In this limit $\chi=\dot{\tilde\varphi}$, therefore $\tilde\varphi$ in the second decomposition becomes equivalent to $\varphi$ in the first one.

The situation is similar in massive gravity. Substituting \eqref{eq:H_pi} in Fierz-Pauli term  one obtains up to quadratic order
\beq
\label{eq:PF_2nd}
&\frac{1}{4}m^2M_{pl}^2(H^2-H_{\mu\nu}^2)=~~~~~~~~~~~~~~~~~~~~~~~~~~~~~~~~~~~~~~~~~~~~~~~~~~~~~~~~~~~~~~~~~~~~~~~~~~~~&\nonumber \\
 &m^2M_{pl}^2\left[\frac{1}{4}(h^2-h_{\mu\nu}^2)
         +h\partial_\mu\pi^\mu-h^{\mu\nu}\partial_\mu\pi_\nu
         -\frac{1}{4}(\partial_\mu\pi_\nu-\partial_\nu\pi_\mu)^2
         +\text{cubic}\right]\,,&
\eeq
where indices are raised by $\eta^{\mu\nu}$. Instead of \eqref{eq:parametrization}, $\pi_\mu$ may be decomposed into 3-scalar and 3-vector parts in the following non-covariant way
\beq
\label{non_cov}
\pi_0=\chi\,,\quad \pi_i = a^T_i+\partial_i\tilde\varphi\,.
\eeq
The part of \eqref{eq:PF_2nd} which contains $\chi$ then reads
\beq
\label{eq:L_chi}
{\cal L}_\chi=m^2M_{pl}^2 \left[\frac{1}{2}(\partial_i\dot{\tilde\varphi}
                     -\partial_i\chi)^2-\dot\chi h+\partial_\mu\chi h^\mu_0
                  +\text{cubic}\right]\,,
\eeq
and varying with respect to $\chi$ one finds
\beq
\label{eq:chi}
\chi=\dot{\tilde\varphi}+\frac{1}{\Delta}(\dot h - \partial_\mu h_0^\mu)
          +\text{quadratic}\,.
\eeq
The appearance of $\dot{\tilde\varphi}$ on the r.h.s. ensures that after substituting \eqref{eq:chi} back in \eqref{eq:PF_2nd} and taking decoupling limit one recovers \eqref{eq:gravity_decouple} for $\tilde\varphi$. To see this more explicitly note that


i) There remains no pure kinetic term for $\tilde\varphi$ (due to cancellation of linear terms in $\tilde\varphi$ in the structure $\partial_i\dot{\tilde\varphi}-\partial_i\chi$ appearing in \eqref{eq:L_chi}) rather it kinetically mixes with $h_{\mu\nu}$ via $\tilde\varphi (\eta_{\mu\nu}\Box-\partial_\mu\partial_\nu)h^{\mu\nu}$. Therefore the canonical normalization for $\tilde\varphi$ is the same as $\varphi$ in \eqref{eq:parametrization}\footnotemark, i.e. $\tilde\varphi^c=\Lambda_3^3\tilde\varphi$. This similarity continues to hold for the vector modes, namely $a_i^c=mM_{pl}a_i$. 

\footnotetext{
There are quadratic corrections to the action, of the form $m^2M_{pl}^2(\partial_\mu h)^2/\Delta$ but they are negligible compared to Einstein-Hilbert action as $m\to 0$. This also implies $h^c\simeq M_{pl}h$.
}

ii) Suppose for the moment that 
\beq
\label{chi}
\chi=\dot{\tilde\varphi}\,.
\eeq
In order to get the most relevant interactions in \eqref{eq:PF_2nd} we can limit ourselves to 
\beq
\label{pi_cube}
m^2M_{pl}^2[\partial_\mu\pi_\mu(\partial_\alpha\pi_\beta)^2-\partial_\mu\pi_\nu\partial_\mu\pi_\alpha\partial_\nu\pi_\alpha]\,,
\eeq
with all indices contracted by $\eta^{\mu\nu}$. The reason is that all other terms either contain powers of $h_{\mu\nu}$ or are of quartic order in $\pi_\mu$, and vanish in the decoupling limit \eqref{Lambda5}. As an example of the first case consider $m^2M_{pl}^2h(\partial_\mu\pi_\mu)^2$, it schematically contains the following terms
\beq
m^2M_{pl}^2 h(\partial a)^2\,,\quad m^2M_{pl}^2h(\partial a\partial^2\tilde\varphi)\,,\quad m^2M_{pl}^2h(\partial^2\tilde\varphi)^2\,,
\eeq
which after canonical normalization are suppressed by $M_{pl}$, $mM_{pl}$ and $m^2M_{pl}$ respectively\footnote{Because of the relative suppression of terms containing $a_i$, we ignore them from now on.}. Quartic interactions, like $m^2M_{pl}^2(\partial_\mu\pi_\mu)^4$, lead to terms that are suppressed at least by $M_{pl}^2m^6$. All these scales go to infinity in decoupling limit \eqref{Lambda5} and the corresponding interactions vanish. Substituting \eqref{chi} in \eqref{pi_cube} and ignoring terms containing $a_i$ (as justified above) is equivalent to writing
\beq
\pi_\mu=\partial_\mu \tilde{\varphi}.
\eeq
Therefore one recovers the cubic part of \eqref{eq:gravity_decouple}.

iii) Because of the cancellation of the linear terms in $\tilde\varphi$, mentioned in (i), the quadratic corrections in \eqref{eq:chi} contribute only to cubic terms of the form $h(\partial^2\tilde\varphi)^2$ and higher order terms. However all of them can be ignored for reasons similar to what was explained in (ii). Consequently the most relevant Lagrangian (the one that survives in decoupling limit) is \eqref{eq:gravity_decouple}.


It is easy to show that the parametrizations \eqref{eq:parametrization} and \eqref{non_cov} are physically equivalent regardless of the form of the potential $V(H_{\mu\nu})$: In Coulomb gauge ($\partial_iA_i=0$), the Lagrangians obtained using \eqref{eq:parametrization} and \eqref{non_cov} are the same, except that $A_0+\dot\varphi$ in the first is replaced by $\chi$ in the second. Thus, integrating out $A_0$ and $\chi$ from corresponding Lagrangians result in identical theories since the solution for $\chi$ is $\dot{\tilde\varphi}$ plus the solution for $A_0$\footnotemark. Therefore, although parametrizations which contain time derivative seem to introduce fake ghosts in the theory, the one caught in \cite{Creminelli,DeffayetRombouts} is not of this kind because \eqref{eq:parametrization} is invariant under a new $U(1)$ gauge symmetry. 

\footnotetext{
With $\varphi$ replaced by $\tilde\varphi$, and $A_i$ by $a_i$.
}


\subsubsection*{Acknowledgments}
We are grateful to Gregory Gabadadze for enlightening discussions. We would also like to thank Viatcheslav Mukhanov for interesting comments. LB is thankful to Gia Dvali for his support through David and Lucile Packard Foundation Fellowship. MM was partially supported by the NSF (grant AST-0908357), NASA (grant NNX08AJ48G), and the NYU James Arthur graduate fellowship.

\subsubsection*{Appendix}

In this section we derive the above mentioned equivalence relation (\ref{sim}). Just to reiterate, the expressions for $H_{\mu \nu}$ and $\bar{h}^{AB}$ are given by
\beq
H_{\mu \nu}=g_{\mu \nu}- \partial_\mu \phi^A \partial_\nu \phi^B \eta_{AB},\\
\bar{h}^{AB}=g^{\mu \nu} \partial_\mu \phi^A \partial_\nu \phi^B-\eta^{AB}.
\eeq
In what follows all repeated space-time indices are contracted by inverse metric $g^{\mu \nu}$, while the field-space ones by $\eta_{AB}$. We also adobt a notation $[\ldots]$ to denote the trace of the tensor. It is easy to check that the following identities hold
\beq
H\equiv H_{\mu \nu} g^{\mu \nu}=-\bar{h}^{AB} \eta_{AB}\equiv -\bar{h},~~~
[H_{\mu \nu}^2]=[\bar{h}_{AB}^2].
\eeq
Taking this into account, we can use induction to prove the identity $[H_{\mu \nu}^{n+1}]=(-1)^{n+1}[\bar{h}_{AB}^{n+1}]$, assuming
\beq
[H_{\mu \nu}^n]=(-1)^n [\bar{h}_{AB}^n].
\label{con}
\eeq
A simple calculation shows that
\beq
[H_{\mu \nu}^{n+1}]&=&[H_{\mu \nu}^n]-H_{\mu \alpha}^{n-1}g^{\mu \nu}g^{\alpha \beta}\partial_\beta \phi^A \partial_\nu \phi^B \bar{h}^{AB}\nonumber \\ ~
&=&[H_{\mu \nu}^n]+H_{\mu \alpha}^{n-2}g^{\mu \nu}g^{\alpha \beta}\partial_\beta \phi^A \partial_\nu \phi^C \bar{h}^{AB} \bar{h}^{BC}=\ldots\nonumber \\  ~
&=&[H_{\mu \nu}^n]+(-1)^{k+1}H_{\mu \alpha}^{n-k}g^{\mu \nu}g^{\alpha \beta}\partial_\beta \phi^A \partial_\nu \phi^B \bar{h}^k _{AB}.
\eeq
From the last equality one obtains $[H_{\mu \nu}^{n+1}]=[H_{\mu \nu}^{n}]-(-1)^n [\bar{h}_{AB}^n]+(-1)^{n+1} [\bar{h}_{AB}^{n+1}]$, which after using (\ref{con}) reduces to (\ref{sim}).

\end {document}